\pdfoutput=1 % to force arxiv to use pdflatex
\documentclass[pra,twocolumn]{revtex4-1}

\usepackage{graphicx,color,transparent}
\usepackage[dvipsnames]{xcolor}
\usepackage{amssymb}
\usepackage{bm}
\usepackage{amsmath,amsfonts,latexsym}

\usepackage{braket} % For bra-ket notation

\usepackage{comment}

\usepackage{bm} % allows bold greek letters

\newcommand*{\cA}{{\cal A}}

\newcommand*{\cD}{{\cal D}}

\newcommand*{\cR}{{\cal R}}

\newcommand*{\cI}{{\cal I}}

\newcommand*{\cS}{{\cal S}}

\begin{document}

\title{Diffractive scattering of three particles in one dimension: a simple result for weak violations of the Yang--Baxter equation}

\author{Austen Lamacraft}
\affiliation{TCM Group, Cavendish Laboratory, University of Cambridge, J. J. Thomson Ave., Cambridge CB3 0HE, UK}
\date{\today}
\email{al200@cam.ac.uk}

\date{\today}

\begin{abstract}
We study scattering of three equal mass particles in one dimension. Integrable interactions are synonymous with \emph{non-diffractive} scattering, meaning that the set of incoming momenta for any scattering event coincides with the set of outgoing momenta. A system is integrable if the two particle scattering matrix obeys the Yang--Baxter equation. Nonintegrable interactions correspond to diffractive scattering, where the set of outgoing momenta may take on all values consistent with energy and momentum conservation. Such processes play a vital role in the kinetics of one dimensional gases, where binary collisions are unable to alter the distribution function.

When integrability is broken weakly, the result is a small diffractive scattering amplitude. Our main result is a simple formula for the diffractive part of the scattering amplitude, when the violation of the Yang--Baxter equation is small. Although the derivation is given for $\delta$-function interactions, the result depends only on the two-particle scattering matrix, and should therefore also apply to finite-range interactions close to integrable. %In comparison, perturbation theory requires the form of the wavefunctions in the near-field region.

% \texttt{Is it true that finite range interactions work? The Sommerfeld rep is presumably only good in the far field anyway -- it solves the wave equation without . Still the nullification theorem should work}
% 
% \texttt{Another result one could hope to find is the form of the wavefunction close to the phase jump -- not doing naive stationary phase.}

\end{abstract}

\maketitle

\section{Introduction}
\label{sec:introduction}

\subsection{Background}
\label{sub:background}

In classical mechanics there is an unambiguous notion of complete integrability, namely that a system with $N$ degrees of freedom must have $N$ (Poisson) commuting constants of motion \cite{arnold1989}. Finding the appropriate analog of this idea in quantum mechanics turns out to be frustratingly difficult: see Ref.~\cite{caux2011} for a recent discussion of the pitfalls encountered. 

In Ref.~\cite{sutherland2004}, Sutherland persuasively advocates a definition of integrability that applies equally well to both classical and quantum systems of \emph{equal mass} particles that can move far away from each other (usually in one dimension), out of the range of any interaction between particles. In this asymptotic region, the momentum of each individual particle is conserved. Scattering processes where the particles begin and end in the asymptotic region are therefore characterized by the initial $\left\{k_{i}\right\}$ and final $\left\{k_{i}'\right\}$ set of momenta. An integrable system is then understood to be one in which these two sets are identical. In the classical case, this means that varying the `impact parameters' for the scattering problem does not change the set of final momenta, while in the quantum case, the scattering amplitude vanishes for all other assignments of momenta to the outgoing particles. Note that in either case, the identity of these two sets does not exclude momenta from being exchanged among the particles undergoing scattering. In analogy with optics, scattering with this character is called \emph{non-diffractive}.

Despite being superficially rather different, it is not hard to see that these two notions of integrability coincide in the classical case. Whatever the general form of the $N$ constants of the motion for the $N$ particles undergoing scattering, they must be equivalent to (i.e. functionally dependent upon) the $N$ momenta in the asymptotic region, which are therefore unchanged after scattering. In quantum mechanics, the \emph{Bethe ansatz} is the non-diffractive form of the $N$-particle wavefunction that forms the foundation for the study of integrable systems in one dimension.

For $N=2$ particles, energy and momentum conservation guarantee $\left\{k_{1},k_{2}\right\}=\left\{k_{1}',k_{2}'\right\}$. The distinction between integrable and nonintegrable systems therefore appears first for $N=3$ particles. For an incoming plane wave $e^{i\left(k_{1}x_{1}+k_{2}x_{2}+k_{3}x_{3}\right)}$, we write the scattered wave in the asymptotic region as
\begin{widetext}
	\begin{multline}
		\label{3ParticlePaper_3partwave}
		\Psi_{3}(x_{1},x_{2},x_{3})\to \sum_{P}\cA_{P}\exp\left[i\left(k_{P1}x_{1}+k_{P2}x_{2}+k_{P3}x_{3}\right)\right]\\
		+\int_{P,E \text{ fixed}}dk_{1}'dk_{2}'dk_{3}'\, \cA_{\text{diff}}(k_{1}',k_{2}',k_{3}')\exp\left[i\left(k_{1}'x_{1}+k_{2}'x_{2}+k_{3}'x_{3}\right)\right].
	\end{multline}
\end{widetext}
The first term involves a sum of the $N!=6$ permutations of the incoming momenta, with an amplitude $\cA_{P}$ for each. This part is the Bethe ansatz wavefunction, and for an integrable system there is nothing more (strictly we must write an expression of this form in each of the six asymptotic regions $x_{Q1}\ll x_{Q2}\ll x_{Q3}$, for all six permutations $Q$, but we do not include this extra detail for now). The second term, appearing only for nonintegrable systems, is the diffracted wave, a superposition of plane waves where the momenta are restricted to a (1D) manifold of fixed total momentum $P=\sum_{i=1}^{3}k_{i}$ and energy $E=\sum_{i=1}^{3}\frac{k_{i}^{2}}{2m}$. The amplitude $\cA_{\text{diff}}(k_{1}',k_{2}',k_{3}')$ is the \emph{diffraction amplitude}.

For a nonintegrable Hamiltonian the three body problem is intractable, even in one dimension. However, a system that is close to integrable, that is, whose Hamiltonian deviates only a little from that of an integrable model, is expected to display a small amount of diffractive scattering. While the primary goal of this work is to obtain the diffraction amplitude for three particles in this limit, we also seek an understanding of \emph{why} diffraction is sometimes absent. In the theory of integrable systems, a distinguished role is played by the \emph{Yang--Baxter equation}, a relation obeyed by the two particle $S$-matrix of an integrable Hamiltonian. In deriving our result, we will see how the violation of the Yang--Baxter equation gives rise to diffraction.

Our result yields valuable insight into how the defining characteristic of integrable systems breaks down upon changing the Hamiltonian. A more practical motivation is provided by recent experiments on one dimensional ultracold atomic gases \cite{kinoshita2006}, showing essentially no relaxation towards an equilibrium state. Recall that in a three dimensional gas, the dominant process of equilibration is binary collisions between gas particles, whose effect on the evolution of the distribution function is described by the Boltzmann equation. In one dimension, such collisions result in $k_{1}=k_{1}'$, $k_{2}=k_{2}'$, or $k_{1}=k_{2}'$, $k_{2}=k_{1}'$, which does not alter the distribution of particles in momentum space and therefore cannot lead to equilibration. Evidently diffractive scattering is required, and three particle collisions will be the most important at low density \cite{mazets2010,tan2010}. We defer to the Conclusion further discussion of kinetics due to three particle collisions.

To what extent are one dimensional atomic gases described by integrable Hamiltonians, or those close to integrable? As shown in Ref.~\cite{olshanii1998}, tight confinement in the two transverse directions allows the interaction between a pair of atoms to be described by a $\delta$-function
\begin{equation}
	\label{3ParticlePaper_1ddelta}
	V(x_{1}-x_{2})=g_{12}\delta(x_{1}-x_{2}).
\end{equation}
A gas of atomic bosons therefore provides (ignoring any confining potential along the length of the gas) a realization of the Lieb--Liniger model of the 1D Bose gas \cite{lieb1963}, soluble by the Bethe ansatz. To destroy integrability one may add external potentials, introduce a more complicated (finite range) interaction between the particles, or consider multiple species. While species with different masses certainly lead to diffractive scattering (as the above considerations should make clear), a more relevant situation in ultracold physics is to consider different internal states of the atoms (e.g. different hyperfine states), in which cases all masses remain identical. However, the interparticle interaction is dependent upon the species involved, and this again leads to a nonintegrable Hamiltonian. Since the variation in the interaction strengths is typically on the order of a few percent, this provides a natural setting for the question of how weak violations of integrability give rise to diffractive scattering.

\subsection{$\delta$-function potentials and the relation to diffraction from a wedge}
\label{sec:deltafunction}

Motivated by the above discussion, we take as our principal example the three particle Hamiltonian 
\begin{multline}
	\label{3Particle_Ham}
	H = p_{1}^{2}+p_{2}^{2}+p_{3}^{2}\\
	+g_{12}\delta(x_{1}-x_{2})+g_{13}\delta(x_{1}-x_{3})+g_{23}\delta(x_{2}-x_{3})
\end{multline}
All masses are equal to $1/2$, and we will assume without loss of generality that all three particles are distinguishable -- the scattering amplitude for the case where two particles are identical bosons or fermions can be constructed later by symmetrizing or antisymmetrizing the solution. The simplicity of Eq.~\eqref{3Particle_Ham} is deceptive: other than the case $g_{12}=g_{13}=g_{23}$ soluble by the Bethe ansatz, only a few other special cases have been solved to date \cite{mcguire1964,mcguire1972,gaudin1975,lipszyc1980}. The general formulation of the problem given in Refs.~\cite{lipszyc1980,mcguire1988} is forbiddingly complex, involving the solution of a system of functional equations (see Eq.~\eqref{3ParticlePaper_AllARel}). In contrast, we are interested in finding the diffraction amplitude when $g_{12}\sim g_{13}\sim g_{23}$. Numerical treatments of various aspects of the scattering problem can be found in Refs.~\cite{mehta2005,mehta2007,kartavtsev2009}.

\begin{figure*}
	\centering
		\includegraphics[width=1.6\columnwidth]{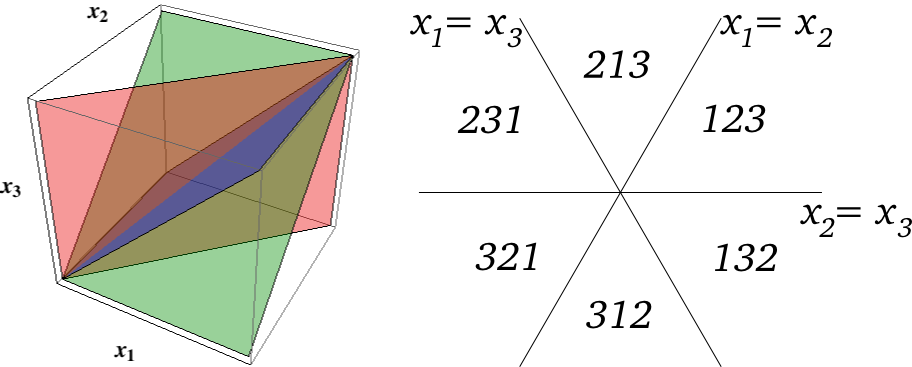}
	\caption{Geometrical description of three particle scattering in real space. (Left) Particles interact on the three planes defined by $x_{i}=x_{j}$. (Right) Projection along the centre of mass motion in the $(1,1,1)$ direction. The six sectors correspond to different ordering of the three particles on the line, given by a three digit code.}
	\label{fig:mirrors}
\end{figure*}

For Eq.~\eqref{3Particle_Ham}, the result \emph{could} be obtained using perturbation theory, starting from the Bethe ansatz form of the wavefunction. However, such an approach makes no explicit connection to the violation of the Yang--Baxter equation, and obscures the fact that the diffraction amplitude can be expressed using only two particle scattering data. This last feature strongly suggests that the result is applicable to \emph{finite range} interactions close to integrable and not just to $\delta$-function interactions. As an example, the interaction potential $V(x_{1}-x_{2})=g_{12}\sinh^{-2}(x_{1}-x_{2})$ is integrable \cite{sutherland2004}), so a small deviation from this potential will introduce a correspondingly small amount of diffractive scattering, described by our main result, Eq.~\eqref{3ParticlePaper_Dsmall}.

The solution of the three body problem is greatly facilitated by the use of the following picture \cite{mcguire1964}. By momentum conservation the wavefunction has only trivial dependence on the $(1,1,1)$ direction in $(x_{1},x_{2},x_{3})$ space, corresponding to center of mass motion, and we may restrict our attention to the plane perpendicular to this direction. In this plane, the surfaces $x_{1}=x_{2}$, $x_{1}=x_{3}$, and $x_{2}=x_{3}$ appear as three lines at angle $\pi/3$ to each other, cutting the plane into six sectors corresponding to the 6 possible permutations of the positions of the particles on the line (Fig.~\ref{fig:mirrors}). The Hamiltonian describing propagation in the plane is just the two dimensional Laplacian, with the $\delta$-function interactions corresponding to boundary conditions at the surfaces. We note that the effect of different masses may be incorporated into this picture by rescaling the spatial coordinates, so that the kinetic energy in the centre of mass frame is isotropic, at the expense of altering the angles between the three planes. %The details of this transformation, and the relationship between the scattering amplitudes in the original and reduced (two dimensional)  problems are given in Appendix \ref{sec:appA}.

\subsection{The Yang--Baxter equation and its geometrical meaning}
\label{sub:YBequation}

Let us try to describe the above two dimensional wave problem using geometrical optics. We first find the transmission and reflection coefficients for each of the three surfaces, which are defined by the following two particle wavefunction
\begin{multline}
	\label{3ParticlePaper_SmatrixDef}
	\Psi_{2}(x_{i},x_{j})=\\\begin{cases}
		e^{i\left(k_{1}x_{i}+k_{2}x_{j}\right)}+r_{ij}(k_{1},k_{2})e^{i(k_{1}x_{j}+k_{2}x_{i})} & x_{i}<x_{j} \\
		t_{ij}(k_{1},k_{2})e^{i(k_{1}x_{i}+k_{2}x_{j})} & x_{i}>x_{j},
	\end{cases}
\end{multline}
A straightforward calculation using the Hamiltonian Eq.~\eqref{3Particle_Ham} yields
\begin{equation}
	\label{3ParticlePaper_Smatrices}
	\begin{split}
		t_{ij}(k_{1},k_{2})=\frac{k_{1}-k_{2}}{k_{1}-k_{2}+ig_{ij}} \\
		r_{ij}(k_{1},k_{2})=\frac{-ig_{ij}}{k_{1}-k_{2}+ig_{ij}}.		
	\end{split}
\end{equation}
Eq.~\eqref{3ParticlePaper_SmatrixDef} describes a collision in which transmission leads to the particle with momentum $k_{1}$ overtaking the particle with momentum $k_{2}$. When tracing rays in the two dimensional picture, then, this means that $k_{1}-k_{2}=2k\sin\alpha>0$, where $0\leq \alpha\leq \pi$ is the angle between the ray and the plane, and $k=\sqrt{E-P^{2}/3}$ is the magnitude of the momentum in the centre of mass frame. We will therefore write the transmission and reflection as a function of $\alpha$ as
\begin{equation}
	\label{3ParticlePaper_Smatricesalpha}
	\begin{split}
		t_{ij}(\alpha)=\frac{2k\sin\alpha}{2k\sin\alpha+ig_{ij}} \\
		r_{ij}(\alpha)=\frac{-ig_{ij}}{2k\sin\alpha+ig_{ij}}.		
	\end{split}
\end{equation}
\begin{figure*}
		\def\svgwidth{0.62\columnwidth}
		(a)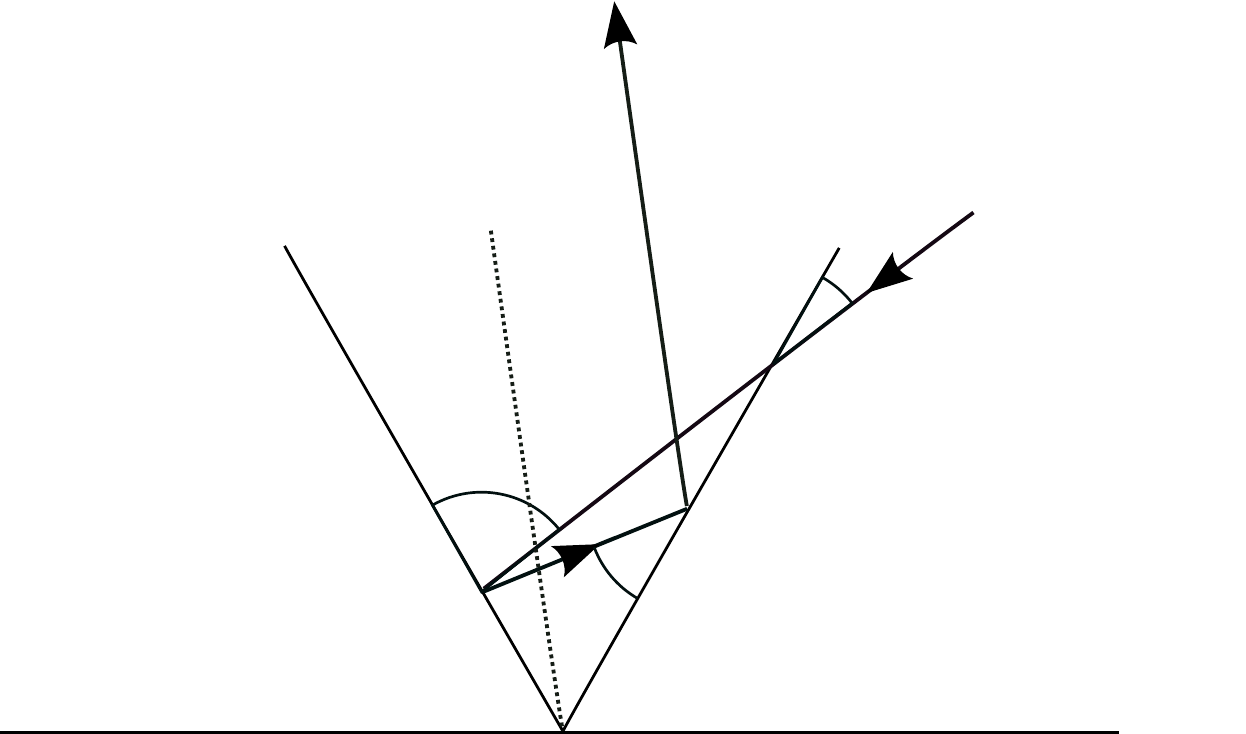
		\def\svgwidth{0.62\columnwidth}
		(b)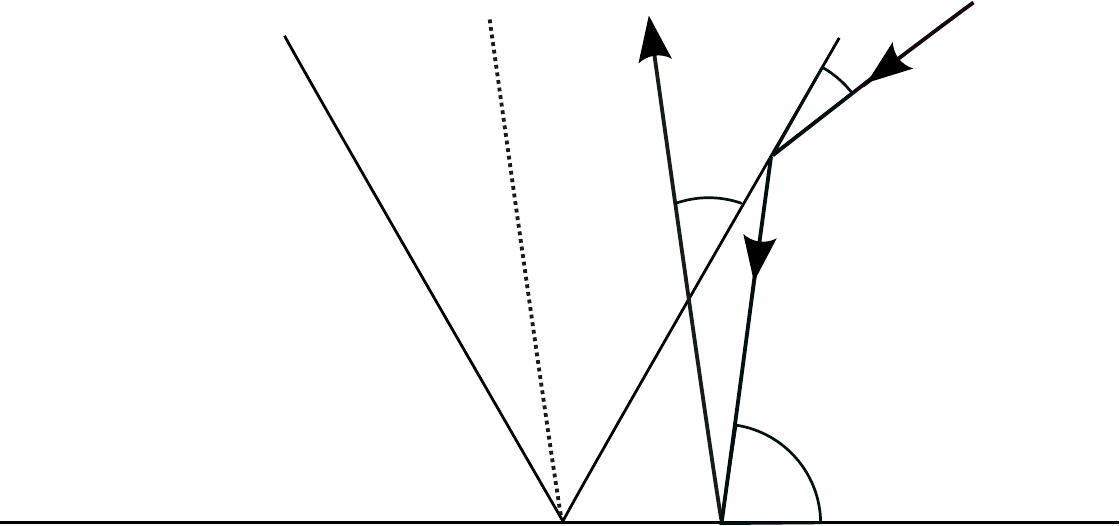
		\def\svgwidth{0.62\columnwidth}
		(c)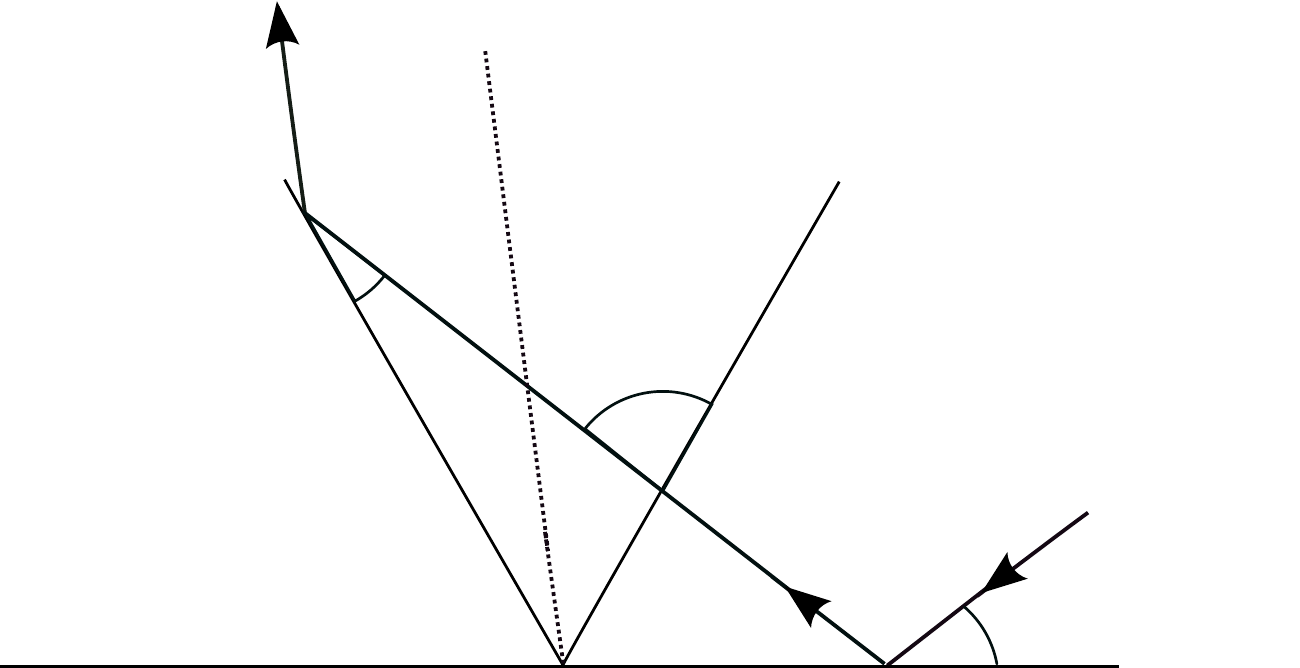
		\caption{Three rays arriving in sector 123 at the same angle, and departing in sector 213, again at the same angle. The paths (a) and (b) contribute to the outgoing wave to the right of the dotted line, while to the left only (c) contributes.
\label{fig:3rays}}
\end{figure*}
With this convention fixed, we consider different ray trajectories with the same arrival and departure angles (Fig.~\ref{fig:3rays}). All three rays illustrated have the same path length, so the difference in their associated amplitudes arises entirely from the reflections and transmissions each experiences. For the outgoing wavefront to the right of the dotted line, two trajectories contribute, depending on whether the wave is first transmitted or reflected, so the overall amplitude is the sum
\begin{multline}
	\label{3ParticlePaper_2rays}
	t_{12}(\alpha)r_{13}(\pi/3+\alpha)r_{12}(\pi/3-\alpha)\\+r_{12}(\alpha)r_{23}(\pi/3+\alpha)t_{12}(\pi/3-\alpha),
\end{multline}
with $0<\alpha<\pi/3$. To the left of the dotted line, only one trajectory contributes, with amplitude
\begin{equation}
	\label{3ParticlePaper_1ray}
	r_{23}(\pi/3-\alpha)t_{12}(\pi/3+\alpha)r_{13}(\alpha).
\end{equation}
\emph{If} the amplitudes in Eq.~\eqref{3ParticlePaper_2rays} and Eq.~\eqref{3ParticlePaper_1ray} are equal, it is plausible that the outgoing wave in sector 213 can be written as a plane wave with amplitude equal to this common value. Returning to the three particle picture: if the incoming wave in sector 123 corresponds to $e^{i(k_{1}x_{1}+k_{2}x_{2}+k_{3}x_{3})}$, the outgoing wave in sector 213 corresponds to $\cA_{231} e^{i(k_{2}x_{1}+k_{3}x_{2}+k_{1}x_{3})}$, where the subscript $231$ on the amplitude indicates how the momenta have been permuted. Comparing with Eq.~\eqref{3ParticlePaper_3partwave}, we see that this corresponds to one term of the Bethe ansatz wavefunction. 

For the outgoing rays corresponding to the other permutations of the momenta, one could draw similar sets of trajectories. Once again, if equality holds between amplitudes for rays contributing to different parts of the outgoing wavefront, it seems plausible -- and we will show explicitly later -- that the Bethe ansatz gives the complete form of the wavefunction. That is, there is no diffraction.

The required equality of Eq.~\eqref{3ParticlePaper_2rays} and Eq.~\eqref{3ParticlePaper_1ray} for arbitrary $\alpha$ is (one component of) the Yang--Baxter equation. Evidently, it is unlikely to be satisfied for an arbitrary set of reflection and transmission coefficients. However, for Eq.~\eqref{3ParticlePaper_Smatricesalpha}, it is satisfied when (and only when) $g_{12}=g_{13}=g_{23}$.

What happens in the general case? In qualitative terms, the wavefronts to the right and left of the dotted line in Fig.~\ref{fig:3rays} will not `match', having different amplitudes. As we move away from the geometrical optics limit, we expect diffraction to smear out this discontinuity, which sheds light on the connection between the violation of the Yang--Baxter equation and the appearance of diffractive scattering. We next turn to the main tool that will be used to make this connection precise. 

%\texttt{Emphasize difference from changing masses.}

% Different masses open up a `shadow region' in the geometric optics limit, as described in Ref.~\cite{sutherland2004} (the earliest reference I could find to the `wedge' picture is Ref.~\cite{mcguire1964}). However for the case of equal masses, but with general interaction parameters between the three particles, nothing seems wrong in the ray optics limit: only three momenta are allowed, and they switch upon reflection. The generation of a continuum of possible momenta upon scattering is thus an intrinsically quantum (wave) effect.

\subsection{Sommerfeld integral}
\label{sec:sommerfeld}

\begin{figure}
	\centering
		\includegraphics[width=\columnwidth]{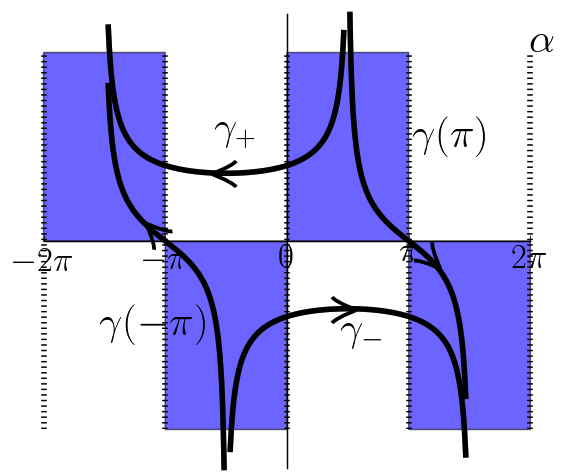}
	\caption{Contour $\gamma=\gamma_{+}\cup\gamma_{-}$ in the $\alpha$ plane for the Sommerfeld integral Eq.~\eqref{3Particle_Sommerfeld} \cite{osipov1999}.}
	\label{fig:Contour}
\end{figure}

When the Bethe ansatz does not work, we need a more general representation of the wavefunction. This is provided by the Sommerfeld integral, which provides a representation of the wavefunction $\Psi_{Q}(r,\phi)$ in sector $Q$, expressed in polar coordinates \cite{sommerfeld2004}
\begin{equation}
	\label{3Particle_Sommerfeld}
	\Psi_{Q}(r,\phi) = \frac{1}{2\pi i} \int_{\gamma} e^{-ikr\cos\alpha} \cA_{Q}(\alpha+\phi)d\alpha.
\end{equation}
It is straightforward to verify that Eq.~\eqref{3Particle_Sommerfeld} satisfies the 2D Helmholtz equation $\left[\nabla^{2}+k^{2}\right]\Psi_{Q}=0$, as long as the integrand vanishes at the endpoints. The contour $\gamma$ must be chosen accordingly. Additionally, we require that there is only an outgoing diffracted wave. The choice shown in Fig.~\ref{fig:Contour} \cite{osipov1999}, has the required properties, as may be seen by writing it as 
\[\int_{\gamma}=\oint_{\gamma_{+}-\gamma(-\pi)+\gamma_{-}-\gamma(\pi)}+\int_{\gamma(-\pi)}+\int_{\gamma(\pi)}.\]
The first integral may be evaluated using the residue theorem, while the second and third pass through the saddle points of the integrand at $\phi=\pm \pi$, and so may be evaluated at large $r$ to give \cite{Norris:1999}
\begin{multline}
	\label{3Particle_SteepestDescent}
	\Psi_{Q}(r,\phi)\underset{r\to\infty}{\longrightarrow}\sum_{i} \text{Res}\,\cA_{Q}(\alpha)|_{\alpha=\phi_{Q}^{(i)}}e^{-ikr\cos\left(\phi^{(i)}_{Q}-\phi\right)}\\+\frac{e^{i\left(kr+\pi/4\right)}}{\sqrt{2\pi kr}}\cD_{Q}(\phi),
\end{multline}
where we have defined the \emph{diffraction amplitude} $\cD_{Q}(\phi)\equiv\cA_{Q}(\phi-\pi)-\cA_{Q}(\phi+\pi)$, and $\left\{\phi_Q^{(i)}\right\}$ give the locations of the poles of $\cA_{Q}(\alpha)$ that are contained within the closed contour (after translation by $-\phi$). Comparison with Eq.~\eqref{3ParticlePaper_3partwave} allows us to identify the first term of Eq.~\eqref{3Particle_SteepestDescent} with the Bethe ansatz contribution, while the second term is the diffracted wave. Note that the saddle point at $\alpha=0$ would give rise to an incoming wave, the reason for its exclusion. 

While one is tempted to think of Eq.~\eqref{3Particle_Sommerfeld} as representing a superposition of plane waves with different wavevectors, with amplitude $\cA_{Q}(\alpha)$ at angle $\alpha+\pi$ \footnote{This slightly frustrating fact is due to the factor $e^{-ikr\cos\alpha}$ in Eq.~\eqref{3Particle_Sommerfeld}, which differs from the more natural choice $e^{ikr\cos\alpha}$, but proves to be far more convenient later}, Eq.~\eqref{3Particle_SteepestDescent} makes it clear that the existence of a diffracted wave is intimately connected with the \emph{absence} of periodicity in $\alpha$. Furthermore, as $\phi$ changes, one pole may move outside of the closed contour, while another a distance $2\pi$ away moves inside. When $\cA_{Q}(\alpha)$ is not periodic the resulting switching of the residues contributing to the first term of Eq.~\eqref{3Particle_SteepestDescent} corresponds to crossing the dotted line in Fig.~\ref{fig:3rays}, where the amplitude within the geometrical optics approximation changes abruptly. The resulting jump in the wave amplitude on crossing this line -- not a true discontinuity but smeared on the scale of the wavelength -- is a distinctive feature of the breaking of integrability in the far field.

We will see in Section~\ref{sub:yang_baxter_equation} that in this language the Yang--Baxter equation appears as a condition for the periodicity of $\cA_{Q}(\alpha)$.

% Thus the correct solution is chosen at the outset, which can be contrasted with Ref.~\cite{mcguire1972} where two functions $G(\theta)$ and $H(\theta)$ are introduced, and the elimination of the incoming wave leads to the relation $G(\theta+\pi)=-H(\pi-\theta)$.

\subsection{Outline of this paper}
\label{sub:outline_of_this_paper}

After this lengthy introduction, let us outline the structure of the remainder of this paper. In the next section we will obtain a system of functional equations obeyed by $\cA_{Q}(\alpha)$ and show how the Yang--Baxter equation corresponds to periodicity of $\cA_{Q}(\alpha)$, as well as finding the explicit form of these functions in the integrable case. This is a vital step in the subsequent derivation of our result for weak violations of the Yang--Baxter equation. In the case of attractive interactions, two and three particle bound states can form. In the integrable case, the collision of a two particle bound state with another particle does not lead to disintegration, even when kinematically allowed. Breaking integrability allows this process to occur, and we find the amplitude for this process. In our Conclusion we discuss the formulation of kinetic theory with three particle collisions.

\section{Derivation of the main result}
\label{sec:deriv}

\subsection{The system of equations for $\cA_{Q}(\alpha)$}
\label{sub:Aequations}

The wavefunctions in the different sectors are subject to the boundary conditions of continuity at the line $x_{i}=x_{j}$ between two sectors and the condition on the normal derivative
\begin{equation}
	\label{3Particle_Jump}
	\frac{\partial \Psi}{\partial n}\bigg|^{+}_{-}=g_{ij}\Psi(x_{i}=x_{j}).
\end{equation}
In polar coordinates this becomes
\begin{equation}
	\label{3Particle_PolarBC}
	\frac{1}{r}\frac{\partial \Psi}{\partial \phi}\bigg|^{+}_{-}=g_{ij}\Psi.
\end{equation}
We are going to substitute this into the Sommerfeld integral representation Eq.~\eqref{3Particle_Sommerfeld}, which we rewrite using the symmetry of the contour as
\begin{multline}
	\label{3Particle_SommerfeldRewrite}
	\Psi_{Q}(r,\phi) = \frac{1}{2\pi i}\int_{\gamma_{+}} e^{-ikr\cos\alpha} \left[\cA_{Q}(\alpha+\phi)\right.\\\left.-\cA_{Q}(\phi-\alpha)\right]d\alpha.
\end{multline}
This allows us to use the \emph{nullification theorem} proved in Ref.~\cite{osipov1999}, which tells us that if an integral of the form $\int_{\gamma_{+}}e^{-ikr\cos\alpha} f(\alpha)\,d\alpha=0$, for $f(\alpha)$ odd (and obeying certain technical conditions at $+i\infty$), then $f(\alpha)=0$. In this way we can convert boundary conditions on $\Psi_{Q}(r,\phi)$ to conditions on $\cA_{Q}(\alpha)$.

% \begin{figure}
% 	\centering
% 		\includegraphics[width=0.4\columnwidth]{chamber.png}
% 	\caption{Standard orientation for a sector.}
% 	\label{fig:chamber}
% \end{figure}

\begin{figure}
	\centering
		\includegraphics[width=0.5\columnwidth]{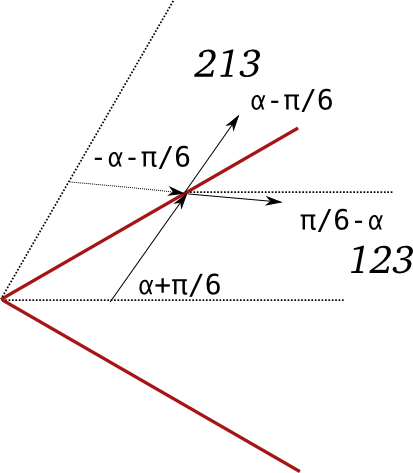}
	\caption{Interpretation of Eq.~\eqref{3Particle_Arelation} in terms of scattering amplitudes. Note that angles are measured from the line bisecting each sector.}
	\label{fig:scatteringsector}
\end{figure}

We orient each sector in a standard way, measuring the angle of our polar coordinates from the bisector of the wedge (Fig.~\ref{fig:scatteringsector}). The above boundary conditions give relations between $\cA_{Q}(\alpha)$ and $\cA_{Q'}(\alpha)$ in neighboring sectors, which we write for the sake of definiteness for sectors 123 and 213 of Fig.~\ref{fig:mirrors}
\begin{widetext}
	\begin{equation}
		\label{3Particle_Arelation}	
		\begin{split}		\cA_{213}(\alpha-\pi/6)=t_{12}(\alpha)\cA_{123}(\alpha+\pi/6)-r_{12}(\alpha)\cA_{213}(-\alpha-\pi/6)\\		\cA_{123}(\pi/6-\alpha)=t_{12}(\alpha)\cA_{213}(-\alpha-\pi/6)-r_{12}(\alpha)\cA_{123}(\alpha+\pi/6)	
		\end{split}		
		\end{equation}	
We can express this equation and the five others arising from the other boundaries in a compact form
\begin{equation}
	\label{3ParticlePaper_AllARel}
	\begin{split}
		\begin{pmatrix}
			-\cA_{123}(\pi/3-\alpha) \\
			\cA_{213}(\alpha-\pi/3) \\
			-\cA_{231}(\pi/3-\alpha) \\
			\cA_{321}(\alpha-\pi/3) \\
			-\cA_{312}(\pi/3-\alpha) \\
			\cA_{132}(\alpha-\pi/3) \\
		\end{pmatrix}=
		\cS_{1}(\alpha-\pi/6)\begin{pmatrix}
		\cA_{123}(\alpha) \\
		-\cA_{213}(-\alpha) \\
		\cA_{231}(\alpha) \\
		-\cA_{321}(-\alpha) \\
		\cA_{312}(\alpha) \\
		-\cA_{132}(-\alpha) \\
	\end{pmatrix},
	\qquad
	\begin{pmatrix}
		-\cA_{123}(\alpha-\pi/3) \\
		\cA_{213}(\pi/3-\alpha) \\
		-\cA_{231}(\alpha-\pi/3) \\
		\cA_{321}(\pi/3-\alpha) \\
		-\cA_{312}(\alpha-\pi/3) \\
		\cA_{132}(\pi/3-\alpha) \\
	\end{pmatrix}=
	\cS_{2}(\alpha-\pi/6)\begin{pmatrix}
		\cA_{123}(-\alpha) \\
		-\cA_{213}(\alpha) \\
		\cA_{231}(-\alpha) \\
		-\cA_{321}(\alpha) \\
		\cA_{312}(-\alpha) \\
		-\cA_{132}(\alpha) \\
	\end{pmatrix}
	\end{split}
\end{equation}
where the matrices $\cS_{1,2}(\alpha)$ are
\begin{align}
	\label{Sdef}
	\cS_1(\alpha)=\begin{pmatrix}
		r_{12}(\alpha) & t_{12}(\alpha) & 0 & 0 & 0 & 0\\
		t_{12}(\alpha) & r_{12}(\alpha) & 0 & 0 & 0 & 0\\
		0 & 0 & r_{23}(\alpha) & t_{23}(\alpha) & 0 & 0\\
		0 & 0 & t_{23}(\alpha) & r_{23}(\alpha) & 0 & 0 \\
		0 & 0 & 0 & 0 & r_{13}(\alpha) & t_{13}(\alpha) \\
		0 & 0 & 0 & 0 & t_{13}(\alpha) & r_{13}(\alpha) 
	\end{pmatrix},\\
	\cS_2(\alpha)=\begin{pmatrix}
		r_{23}(\alpha) & 0 & 0 & 0 & 0 & t_{23}(\alpha)\\
		0 & r_{13}(\alpha) & t_{13}(\alpha) & 0 & 0 & 0\\
		0 & t_{13}(\alpha) & r_{13}(\alpha) & 0 & 0 & 0\\
		0 & 0 & 0 & r_{12}(\alpha) & t_{12}(\alpha) & 0 \\
		0 & 0 & 0 & t_{12}(\alpha) & r_{12}(\alpha) & 0 \\
		t_{23}(\alpha) & 0 & 0 & 0 & 0 & r_{23}(\alpha).
	\end{pmatrix}\\
\end{align}
\end{widetext}
Eq.~\eqref{3ParticlePaper_AllARel} have a natural interpretation in terms of reflection and transmission (Fig.~\ref{fig:scatteringsector}) \footnote{The minus signs in Eq.~\eqref{3ParticlePaper_AllARel} may be at first surprising: in Section~\ref{sub:yang_baxter_equation} we will see that it is more correctly the \emph{residues} of $\cA_{Q}(\alpha)$ that are related by the S-matrix}. They relate $\cA_{Q}(\alpha)$ at an infinite discrete set of $\alpha$ values, which correspond to six different rays (Fig.~\ref{fig:allrays}). Each ray corresponds to an infinite set of amplitudes with $\alpha$ differing by multiples of $2\pi$.

Eqs.~\eqref{3ParticlePaper_AllARel} and \eqref{Sdef} are written in the \emph{reflection diagonal representation}, as the entries of our vectors of amplitudes always correspond to the same sectors. In the \emph{transmission diagonal representation} the entries would correspond to the rays of Fig.~\ref{fig:allrays}.

% Each ray is connected (that is, may be accessed by reflection from) only 3 others, forming the complete bipartite graph $K_{3,3}$ (Fig.~\ref{fig:Biclique_K_3_3})
% %
% \begin{figure}
% 	\centering
% 		\includegraphics[width=0.4\columnwidth]{Biclique_K_3_3.png}
% 	\caption{Connectivity of the six rays, forming the graph $K_{3,3}$}
% 	\label{fig:Biclique_K_3_3}
% \end{figure}
% %
% If we want to get (say) from an incoming pink ray to a green one in a neighboring sector, there are three possibilities (Fig.~\ref{fig:pinktogreen}). 
% 
% \begin{enumerate}
% 	\item 	Transmit (past a blue), reflect to a cyan, then reflect to a green.
% 	\item 	Reflect to a blue, reflect to a green, then transmit (past a cyan).
% 	\item   Reflect to a yellow, transmit (past a red), then reflect to a green.
% \end{enumerate}

\subsection{Yang--Baxter equation and $\cA_{Q}(\alpha)$ in the integrable case}
\label{sub:yang_baxter_equation}

\begin{figure*}
	\centering
		(a)\def\svgwidth{0.6\columnwidth}
		\centering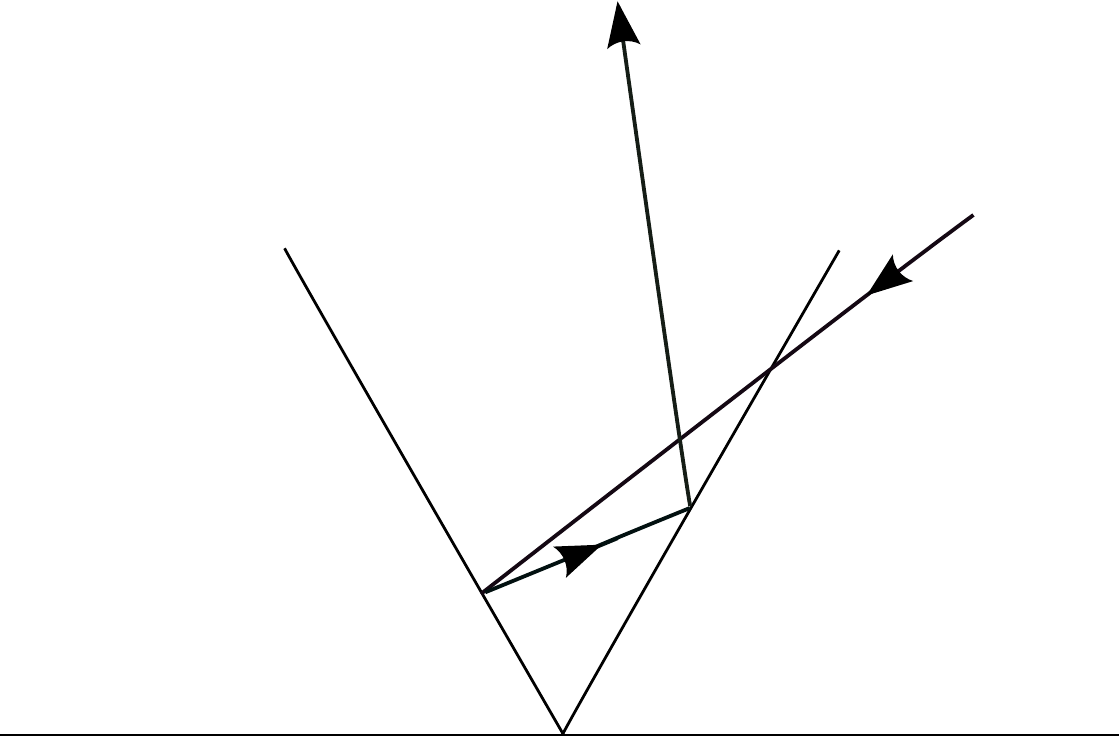
		(b)\def\svgwidth{0.6\columnwidth}
		\centering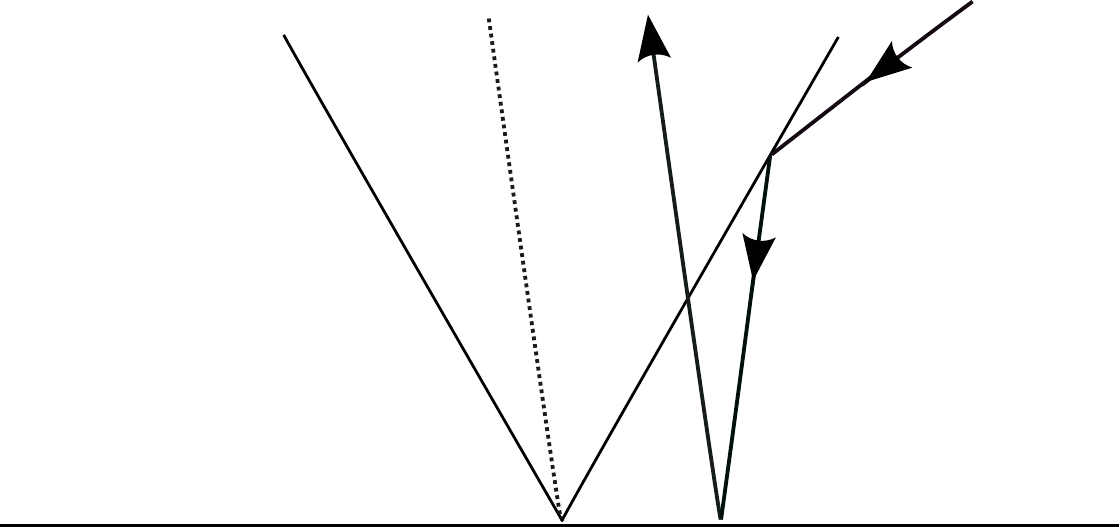
		(c)\def\svgwidth{0.6\columnwidth}
		\centering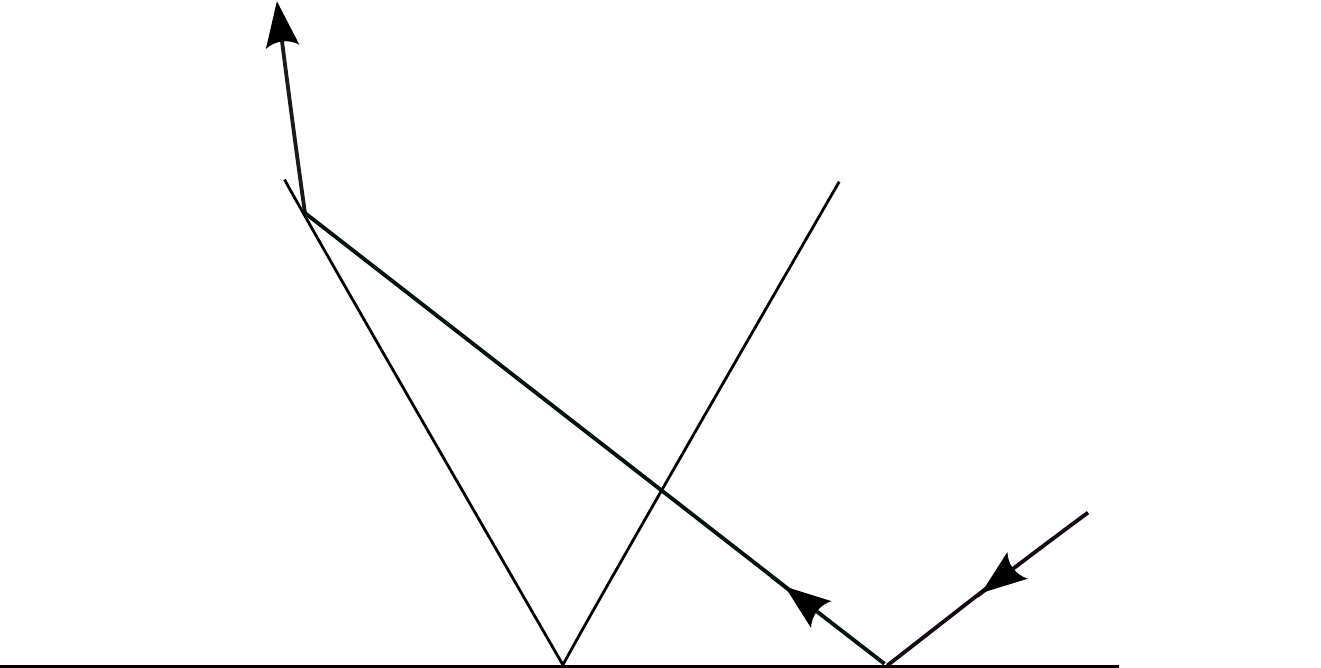		
	\caption{The three rays considered in Section~\ref{sub:YBequation}, now labelled with the arguments of the amplitude $\cA_{Q}(\alpha)$ describing each.}
	\label{fig:pinktogreen}
\end{figure*}

We are now in a position to see how our earlier informal discussion of the Yang--Baxter equation (Section~\ref{sub:YBequation}) reappears in this formalism. By keeping track of the arguments of the $\cA_{Q}(\alpha)$ that appear in Eq.~\eqref{3ParticlePaper_AllARel} and the other five relations, we can trace the amplitude corresponding to each of the rays in Fig.~\ref{fig:pinktogreen}, the same rays we traced before. We see that in the first two cases the final ray amplitude is $\cA_{213}(\alpha-5\pi/6)$, while in the third it is $\cA_{213}(\alpha+7\pi/6)$. As we saw in Section~\ref{sub:YBequation}, the combined amplitude for the first two diagrams is 
\begin{multline}
	\label{3ParticlePaper_2raysagain}
	t_{12}(\alpha)r_{13}(\pi/3+\alpha)r_{12}(\pi/3-\alpha)\\+r_{12}(\alpha)r_{23}(\pi/3+\alpha)t_{12}(\pi/3-\alpha),
\end{multline}
while for the third it is
\begin{equation}
	\label{3ParticlePaper_1rayagain}
	r_{23}(\pi/3-\alpha)t_{12}(\pi/3+\alpha)r_{13}(\alpha).
\end{equation}
Equality of these two amplitudes is then a necessary condition for $\cA_{213}(\alpha-5\pi/6) =\cA_{213}(\alpha+7\pi/6)$ i.e. periodicity of $\cA_{Q}(\alpha)$.

\begin{figure}
	\centering
	\def\svgwidth{1\columnwidth}
	\centering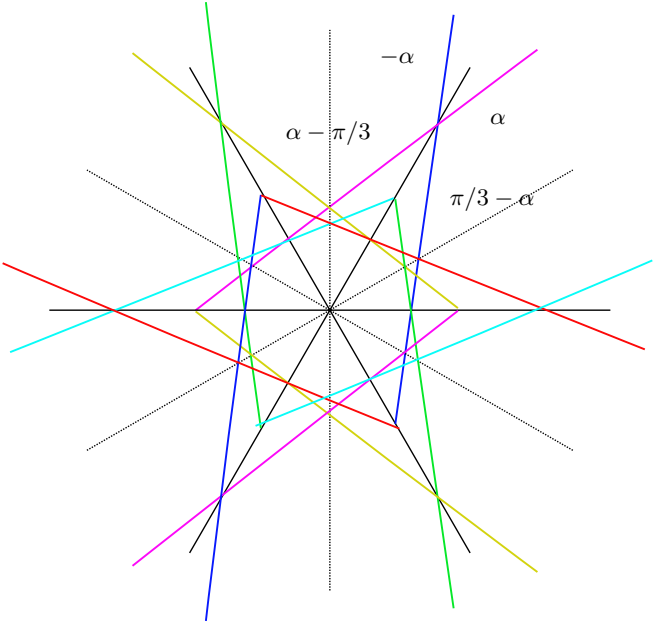
	\caption{All the rays generated by scattering. The color coding makes it clear that each of the six directions is present in each sector.}
	\label{fig:allrays}
\end{figure}

To determine that these relations are sufficient seems daunting at first, as we have to keep track of six sets of amplitudes in six different sectors. However, inspection of the rays in Fig.~\ref{fig:allrays} shows that scattering follows a well-defined order, which provides a great simplification.

Each boundary between sectors corresponding to a collision of two particles has three vertices on it, corresponding to three different scattering events. Scattering proceeds in the same order at each boundary. Thus we only need to keep track of six amplitudes at a time, which are mapped by the S-matrix into 6 others. After scattering three times, the rays are in the sector opposite to where they started. Scattering three times more will bring them back to their original sectors. % Alternatively, we can use parity with respect to transformations $\mathbf{x}\to -\mathbf{x}$ to conclude that the amplitudes may be taken to be $\pm 1$ the amplitudes after scattering three times. 

These six scatterings map the amplitudes as follows
\begin{widetext}
\begin{multline}
	\label{3ParticlePaper_Achain}
	\begin{pmatrix}
		\cA_{123}(\alpha) \\
		-\cA_{213}(-\alpha) \\
		\cA_{231}(\alpha) \\
		-\cA_{321}(-\alpha) \\
		\cA_{312}(\alpha) \\
		-\cA_{132}(-\alpha) \\
	\end{pmatrix}
	\overset{\cS_{1}(\alpha-\pi/6)}{\longrightarrow}
	\begin{pmatrix}
		-\cA_{123}(\pi/3-\alpha) \\
		\cA_{213}(\alpha-\pi/3) \\
		-\cA_{231}(\pi/3-\alpha) \\
		\cA_{321}(\alpha-\pi/3) \\
		-\cA_{312}(\pi/3-\alpha) \\
		\cA_{132}(\alpha-\pi/3) \\
	\end{pmatrix}
	\overset{\cS_{2}(\alpha-\pi/2)}{\longrightarrow}
	\begin{pmatrix}
		\cA_{123}(\alpha-2\pi/3) \\
		-\cA_{213}(2\pi/3-\alpha) \\
		\cA_{231}(\alpha-2\pi/3) \\
		-\cA_{321}(2\pi/3-\alpha) \\
		\cA_{312}(\alpha-2\pi/3) \\
		-\cA_{132}(2\pi/3-\alpha) \\
	\end{pmatrix}
	\overset{\cS_{1}(\alpha-5\pi/6)}{\longrightarrow}
	\begin{pmatrix}
		-\cA_{123}(\pi-\alpha) \\
		\cA_{213}(\alpha-\pi) \\
		-\cA_{231}(\pi-\alpha) \\
		\cA_{321}(\alpha-\pi) \\
		-\cA_{312}(\pi-\alpha) \\
		\cA_{132}(\alpha-\pi) \\
	\end{pmatrix}\\
	\overset{\cS_{2}(\alpha-7\pi/6)}{\longrightarrow}
	\begin{pmatrix}
		\cA_{123}(\alpha-4\pi/3) \\
		-\cA_{213}(4\pi/3-\alpha) \\
		\cA_{231}(\alpha-4\pi/3) \\
		-\cA_{321}(4\pi/3-\alpha) \\
		\cA_{312}(\alpha-4\pi/3) \\
		-\cA_{132}(4\pi/3-\alpha) \\
	\end{pmatrix}
	\overset{\cS_{1}(\alpha-3\pi/2)}{\longrightarrow}
	\begin{pmatrix}
		-\cA_{123}(5\pi/3-\alpha) \\
		\cA_{213}(\alpha-5\pi/3) \\
		-\cA_{231}(5\pi/3-\alpha) \\
		\cA_{321}(\alpha-5\pi/3) \\
		-\cA_{312}(5\pi/3-\alpha) \\
		\cA_{132}(\alpha-5\pi/3) \\
	\end{pmatrix}
	\overset{\cS_{2}(\alpha-11\pi/6)}{\longrightarrow}
	\begin{pmatrix}
		\cA_{123}(\alpha-2\pi) \\
		-\cA_{213}(2\pi-\alpha) \\
		\cA_{231}(\alpha-2\pi) \\
		-\cA_{321}(2\pi-\alpha) \\
		\cA_{312}(\alpha-2\pi) \\
		-\cA_{132}(2\pi-\alpha) \\
	\end{pmatrix},
\end{multline}
Eq.~\eqref{3ParticlePaper_Achain} shows that periodicity of the $\cA_{Q}(\alpha)$ is guaranteed if (it's convenient to shift $\alpha\to \alpha+\pi/6$)
\begin{equation}
	\label{3ParticlePaper_Scondition}
	\cS_{2}(\alpha-5\pi/3)\cS_{1}(\alpha-4\pi/3)\cS_{2}(\alpha-\pi)\cS_{1}(\alpha-2\pi/3)\cS_{2}(\alpha-\pi/3)\cS_{1}(\alpha)=\openone
\end{equation}
Noting that $\cS_{1,2}(\alpha)$ is $2\pi$-periodic, and $\cS_{1,2}(\alpha^{*}+\pi)=\cS_{1,2}^{-1}(\alpha)$, this is equivalent to
\begin{equation}
	\label{3ParticlePaper_YBmatrix}
	\cS_{1}(\alpha-2\pi/3)\cS_{2}(\alpha-\pi/3)\cS_{1}(\alpha)=\cS_{2}(\alpha)\cS_{1}(\alpha-\pi/3)\cS_{2}(\alpha-2\pi/3),
\end{equation}
\end{widetext}
which is the \emph{Yang--Baxter equation}. Parts of this equation are trivial. For example, one can stay in the same sector only by undergoing three reflections, so the diagonal elements of the equation are always satisfied. Likewise one can only get to the opposite sector with three transmissions. The nontrivial elements are those that connect neighboring sectors, and these correspond to the equality of Eq.~\eqref{3ParticlePaper_2raysagain} and Eq.~\eqref{3ParticlePaper_1rayagain} (and the corresponding relations for other neighboring sectors)

Verifying that the matrices given by Eq.~\eqref{Sdef} satisfy Eq.~\eqref{3ParticlePaper_YBmatrix} for the case $g_{12}=g_{13}=g_{23}$ is now a straightforward (if lengthy) exercise.

Now that we have a solution of the Yang--Baxter relations, we still need to find an explicit form for $\cA_{Q}(\alpha)$. From the discussion of Section~\ref{sec:sommerfeld}, we know that the Bethe ansatz form arises solely from the pole contributions to the Sommerfeld integral. Thus we seek functions with the following properties:

\begin{enumerate}
	\item $\cA_{Q}(\alpha)$ are analytic and periodic with period $2\pi$.
	\item Simple poles located at angles corresponding to the direction of the incoming wave, and the five other angles connecting by scattering (Note that there may be other poles off the real $\alpha$ axis coming from the scattering matrix, will will play a role when we consider attractive interactions).
	\item Residues of the poles related by the equations of Section~\ref{sub:Aequations} (Eq.~\eqref{3ParticlePaper_AllARel}). 
\end{enumerate}

Why is a relation between the residues enough to guarantee that Eq.~\eqref{3ParticlePaper_AllARel} is satisfied for all $\alpha$? We know that in the periodic case the Sommerfeld integral is given only by its residues. Thus we can invoke the nullification theorem once more to argue that if the residues have been chosen correctly, the solution must be correct.

To take a simple example, consider the impenetrable case $g_{ij}\to\infty$. Then all reflection amplitudes $r_{ij}=-1$. Fixing the sector to be 123, we seek a function with a pole at $\alpha=\phi_{0}$ with unit residue, corresponding to an incoming wave at angle $\phi_{0}+\pi$. Using Eq.~\eqref{3ParticlePaper_AllARel}, we see that there should be poles with residue $+1$ at $\alpha=\phi_{0}-2\pi/3 \text{ and }\phi_{0}-4\pi/3$, and all angles differing by multiples of $2\pi$, and poles with residue $-1$ at $\alpha=\pi/3-\phi_{0},\,\pi-\phi_{0}, \text{ and }5\pi/3-\phi_{0}$, and all angles differing by multiples of $2\pi$.

One way to construct an analytic function with the correct properties is via the function $p(\alpha)\equiv\frac{1}{2}\cot(\alpha/2)=\sum_{n=-\infty}^{\infty}\frac{1}{\alpha+2\pi n}$ having a set of poles with unit residue at $\alpha=2\pi n$ for integer $n$. Then the function
\begin{multline}
	\label{3ParticlePaper_psum}
	\cA^{\text{imp}}_{123}(\alpha)=p(\alpha-\phi_{0})-p(\alpha-\pi/3+\phi_{0})+p(\alpha+2\pi/3-\phi_{0})\\-p(\alpha-\pi+\phi_{0})+p(\alpha+4\pi/3-\phi_{0})-p(\alpha+\phi_{0}-5\pi/3)
\end{multline}
has the correct poles and residues. Simplifying gives
\begin{equation}
	\label{3ParticlePaper_Aimpsimp}
	\cA_{123}^{\text{imp}}(\alpha)\equiv\frac{3\cos 3\phi_{0}}{\sin 3\alpha-\sin 3\phi_{0}},
\end{equation}
which one may verify has the desired properties. This result was originally obtained by Sommerfeld \cite{sommerfeld2004}, and it should be clear that it works for arbitrary wedge angles $\Phi$ by the replacement $3\to \pi/\Phi$ (with diffraction occurring when $\pi/\Phi$ is non-integer, so that $\cA_{\text{imp}}(\alpha)$ is not periodic in $2\pi$).

The extension to the general case soluble by the Bethe ansatz should now be clear. The residues are related as to one another as implied by Eq.~\eqref{3ParticlePaper_AllARel}. Denoting the residue by $\cR_{Q}(\alpha)\equiv \text{Res}\,\cA_{Q}(\alpha)$ we have, for example
\begin{equation}
	\label{3ParticlePaper_ResRel}
	\begin{pmatrix}
		\cR_{123}(\pi/3-\phi_{0}) \\
		\cR_{213}(\phi_{0}-\pi/3) \\
		\cR_{231}(\pi/3-\phi_{0}) \\
		\cR_{321}(\phi_{0}-\pi/3) \\
		\cR_{312}(\pi/3-\phi_{0}) \\
		\cR_{132}(\phi_{0}-\pi/3) \\
	\end{pmatrix}=
	\cS_{1}(\phi_{0}-\pi/6)\begin{pmatrix}
		\cR_{123}(\phi_{0}) \\
		\cR_{213}(-\phi_{0}) \\
		\cR_{231}(\phi_{0}) \\
		\cR_{321}(-\phi_{0}) \\
		\cR_{312}(\phi_{0}) \\
		\cR_{132}(-\phi_{0}) \\
	\end{pmatrix},
\end{equation}
(Note that the minus signs have disappeared!) and similarly for the $\cS_{2}$ equation, after which one constructs the $2\pi$-periodic analytic function
\begin{widetext}
\begin{multline}
	\label{3ParticlePaper_Aanalytic}
	\cA^{\text{B}}_{Q}(\alpha)\equiv
	\begin{cases}
		\cR_{Q}(\phi_{0}) p(\alpha-\phi_{0})+\cR_{Q}(\pi/3-\phi_{0})p(\alpha-\pi/3+\phi_{0})\\+\cR_{Q}(\phi_{0}-2\pi/3)p(\alpha+2\pi/3-\phi_{0})+\cR_{Q}(\pi-\phi_{0})p(\alpha-\pi+\phi_{0})\\+\cR_{Q}(\phi_{0}-4\pi/3)p(\alpha+4\pi/3-\phi_{0})+\cR_{Q}(5\pi/3-\phi_{0})p(\alpha+\phi_{0}-5\pi/3) & Q=123,231,312 \\
		\cR_{Q}(-\phi_{0}) p(\alpha+\phi_{0})+\cR_{Q}(-\pi/3+\phi_{0})p(\alpha+\pi/3-\phi_{0})\\+\cR_{Q}(-\phi_{0}+2\pi/3)p(\alpha-2\pi/3+\phi_{0})+\cR_{Q}(-\pi+\phi_{0})p(\alpha+\pi-\phi_{0})\\+\cR_{Q}(-\phi_{0}+4\pi/3)p(\alpha-4\pi/3+\phi_{0})+\cR_{Q}(-5\pi/3+\phi_{0})p(\alpha-\phi_{0}+5\pi/3) & Q=213,321,132. \\
	\end{cases}
\end{multline}
\end{widetext}
(the B is for Bethe) As we argued above, Eq.~\eqref{3ParticlePaper_Aanalytic} must satisfy Eq.~\eqref{3ParticlePaper_AllARel} at all $\alpha$. Additionally, we have verified this using Mathematica.

%\texttt{Comment on absence of poles off axis, despite their presence in the reflection and transmission amplitude}

By construction, the Sommerfeld integral of Eq.~\eqref{3ParticlePaper_Aanalytic} reproduces the Bethe ansatz solution with no diffracted wave. To describe an incoming wave at angle $\phi_{0}+\pi$ in the 123 sector we choose
\begin{equation}
	\label{3ParticlePaper_OneWave}
	\begin{pmatrix}
		\cR_{123}(\phi_{0}) \\
		\cR_{213}(-\phi_{0}) \\
		\cR_{231}(\phi_{0}) \\
		\cR_{321}(-\phi_{0}) \\
		\cR_{312}(\phi_{0}) \\
		\cR_{132}(-\phi_{0}) \\
	\end{pmatrix}=
	\begin{pmatrix}
		1 \\ 0 \\ 0\\0 \\0\\0
	\end{pmatrix}
\end{equation}
To deal with the case where two the particles are identical bosons or fermions requires an incoming wave of appropriate symmetry.

% \texttt{Need to invoke something here. Mittag--Leffler says function is unique meromorphic function with these properties}
% 
% \texttt{There are no additional poles coming from the scattering matrix in here!}
% 
% \texttt{Another possibility: If not unique a solution take difference from solution and plug into Sommerfeld integral. According to nullification theorem, it must be the same! Because we deal with periodic functions, we can evaluate Sommerfeld integral exactly (not just in far field)}

\subsection{Diffractive scattering for weak violations of the Yang--Baxter equation}
\label{sub:diffractive_scattering_for_weak_violations_of_the_yang_baxter_equation}

We now move on to the case where the Yang--Baxter equation Eq.~\eqref{3ParticlePaper_YBmatrix} is not satisfied. This problem is much more difficult: we must solve the functional relations Eq.~\eqref{3ParticlePaper_AllARel} without the condition of periodicity in $\alpha$. For this reason, only a few special cases have been worked out in detail \cite{mcguire1964,mcguire1972,gaudin1975,lipszyc1980}.

However, there is a simple result that we can write down with little effort in the case that the Yang--Baxter equation is \emph{nearly} satisfied. Recall that the diffracted wave has the form
\begin{equation}
	\label{3ParticlePaper_DiffWave}
	\frac{e^{i\left(kr+\pi/4\right)}}{\sqrt{2\pi kr}}\cD_{Q}(\phi),
\end{equation}
where $\cD_{Q}(\phi)\equiv\cA_{Q}(\phi-\pi)-\cA_{Q}(\phi+\pi)$. This quantity is small, because $\cA_{Q}(\alpha)$ is almost periodic. Using Eq.~\eqref{3ParticlePaper_Achain}, we have that
\begin{widetext}
\begin{equation}
	\label{3ParticlePaper_3shift}
	\begin{pmatrix}
		-\cA_{123}(\pi-\phi) \\
		\cA_{213}(\phi-\pi) \\
		-\cA_{231}(\pi-\phi) \\
		\cA_{321}(\phi-\pi) \\
		-\cA_{312}(\pi-\phi) \\
		\cA_{132}(\phi-\pi) \\
	\end{pmatrix}=
	\cS_{1}(\phi-2\pi/3)\cS_{2}(\phi-\pi/3)\cS_{1}(\phi)
	\begin{pmatrix}
		\cA_{123}(\phi) \\
		-\cA_{213}(-\phi) \\
		\cA_{231}(\phi) \\
		-\cA_{321}(-\phi) \\
		\cA_{312}(\phi) \\
		-\cA_{132}(-\phi) \\
	\end{pmatrix}
\end{equation}
and
\begin{equation}
	\label{3ParticlePaper_3shift}
	\begin{pmatrix}
		-\cA_{123}(-\pi-\phi) \\
		\cA_{213}(\phi+\pi) \\
		-\cA_{231}(-\pi-\phi) \\
		\cA_{321}(\phi+\pi) \\
		-\cA_{312}(-\pi-\phi) \\
		\cA_{132}(\phi+\pi) \\
	\end{pmatrix}=
	\cS_{2}(\phi)\cS_{1}(\phi-\pi/3)\cS_{2}(\phi-2\pi/3)
	\begin{pmatrix}
		\cA_{123}(\phi) \\
		-\cA_{213}(-\phi) \\
		\cA_{231}(\phi) \\
		-\cA_{321}(-\phi) \\
		\cA_{312}(\phi) \\
		-\cA_{132}(-\phi) \\
	\end{pmatrix}
\end{equation}
are almost equal. In the first approximation, then, we can take
\begin{equation}
	\label{3ParticlePaper_Dsmall}
	\begin{pmatrix}
	\cD_{123}(-\phi) \\
	\cD_{213}(\phi) \\	
	\cD_{231}(-\phi) \\
	\cD_{321}(\phi) \\
	\cD_{312}(-\phi) \\
	\cD_{132}(\phi) \\
	\end{pmatrix}=\left[\cS_{1}(\phi-2\pi/3)\cS_{2}(\phi-\pi/3)\cS_{1}(\phi)-\cS_{2}(\phi)\cS_{1}(\phi-\pi/3)\cS_{2}(\phi-2\pi/3)\right]
	\begin{pmatrix}
	\cA^{\text{B}}_{123}(\phi) \\
	\cA^{\text{B}}_{213}(-\phi) \\	
	\cA^{\text{B}}_{231}(\phi) \\
	\cA^{\text{B}}_{321}(-\phi) \\
	\cA^{\text{B}}_{312}(\phi) \\
	\cA^{\text{B}}_{132}(-\phi) \\
	\end{pmatrix},
\end{equation}
\end{widetext}
where we have substituted the Bethe ansatz form of $\cA_{Q}(\alpha)$. 

Eq.~\eqref{3ParticlePaper_Dsmall} is the main result of this paper. It gives a compact expression for the amplitude of the diffracted wave, valid when the quantity in square brackets (which measures the violation of the Yang--Baxter equation) is small. Substituting the transmission and reflection amplitudes Eq.~\eqref{3ParticlePaper_Smatricesalpha}, one finds that this quantity is first order in the deviation of the interaction constants from the integrable point $g_{12}=g_{13}=g_{23}\neq 0,\infty$. Starting from zero interaction the diffraction amplitude is bilinear in the interaction constants, consistent with the perturbation theory of Ref.~\cite{lunde2007}. For interaction constants close to infinite, the diffraction amplitude is bilinear in $1/g_{ij}$. This is to be expected since the interaction potential $g_{ij}\delta(x_{i}-x_{j})$ may be replaced with $-\left(1/g_{ij}\right)\delta''(x_{1}-x_{2})$ for wavefunctions that vanish at coincident points. 

%\texttt{Features of the scattering amplitude -- glancing blows (two particles leave with close to same momentum)}

%\texttt{Add up poles approach?? Why doesn't it work? One answer would be that although it may be correct in the far field, that isn't enough to be able to prove the nullificaiton theorem. On the other hand Mittag--Leffler may be enough to reconstruct.}

\subsection{Attractive interactions: scattering to and from bound states}
\label{sub:attractive}

With attractive interactions two and three particle bound states appear. A two particle bound state appears as a surface wave on one of the boundaries between sectors. Motion along the boundary corresponds to relative motion of the bound pair and unbound particle. A surface wave corresponds to a pole in $\cA_{Q}(\alpha)$ located at complex $\phi_{Q}^{(i)}$. Bearing in mind Eq.~\eqref{3Particle_SteepestDescent} we have
\begin{equation}
	\label{3ParticlePaper_SurfaceWaveCases}
	\phi_{Q}^{(i)}=\begin{cases}
		\pi/6+i\varphi & \text{ incoming at angle } \phi=\pi/6\\
		-\pi/6-i\varphi & \text{ incoming at angle } \phi=-\pi/6\\
		7\pi/6-i\varphi & \text{ outgoing at angle } \phi=\pi/6\\
		5\pi/6+i\varphi & \text{ outgoing at angle } \phi=-\pi/6\\
	\end{cases}
\end{equation}
where $\varphi>0$, and the signs are chosen in order that the wave decays as we move away from the boundary. For instance, an incoming wave at angle $\phi=\pi/6$ has the form
\begin{multline}\label{boudpairwave}
	e^{-ikr\cos(\pi/6+i\varphi-\phi)}=e^{-ikr\cosh\varphi\cos(\phi-\pi/6)}\\
	\times e^{-kr\sinh\varphi\sin(\pi/6-\phi)},
\end{multline}
which describes a wave with wavevector having components $k\cosh\varphi$ inwards along the boundary, and $ik\sinh\varphi$ perpendicular to it. The second factor is nothing but the two body bound state wavefunction, which allows the identification $\varphi=-\mathop{\text{arcsinh}}(g/2k)$.

In the integrable case inelastic processes in which a bound pair forms or disintegrates cannot occur. Breaking integrability leads to a non-zero amplitude for such processes. To demonstrate these facts requires that we first find the analog of the Bethe solution Eq.~\eqref{3ParticlePaper_Aanalytic} for the motion of bound pairs. 

If we start from an incoming surface wave at $\phi=\pi/6$ in the 123 sector, Eq.~\eqref{3ParticlePaper_Achain} maps the amplitude within this sector at the following arguments
\begin{multline}
	\label{3ParticlePaper_WaveChain}
	\underline{\pi/6+i\varphi}\to \pi/6-i\varphi\to -\pi/2+i\varphi \\
	\to 5\pi/6-i\varphi\to \underline{-7\pi/6+i\varphi}\to \underline{3\pi/2-i\varphi},
\end{multline}
before we repeat ourselves (due to periodicity of $\cA_{Q}(\alpha)$). Inspection of the imaginary parts of the waves reveals that only the underlined amplitudes correspond to waves that decay appropriately at large $r$. Thus there must be no residue at the other values. 

To check that this is so, we start with 
\begin{equation}
	\label{3ParticlePaper_SurfaceWaveIn}
	\begin{pmatrix}
		\cR_{123}(\pi/6+i\varphi) \\
		\cR_{213}(-\pi/6-i\varphi) \\
		\cR_{231}(\pi/6+i\varphi) \\
		\cR_{321}(-\pi/6-i\varphi) \\
		\cR_{312}(\pi/6+i\varphi) \\
		\cR_{132}(-\pi/6-i\varphi) \\
	\end{pmatrix}=
	\begin{pmatrix}
		1 \\ 1 \\ 0\\0 \\0\\0
	\end{pmatrix},
\end{equation}
corresponding to an incoming bound state of particles 1 and 2 at the boundary of sectors 123 and 213. We immediately run into a problem if we try and map forwards through the chain of Eq.~\eqref{3ParticlePaper_Achain}, because $S_{1}(i\varphi)$ is evaluated at the pole corresponding to the two body bound state. So instead we map \emph{backwards}, e.g.
\begin{equation}
	\label{3ParticlePaper_BackChain}
	\begin{pmatrix}
		\cR_{123}(\pi/6+i\varphi) \\
		\cR_{213}(-\pi/6-i\varphi) \\
		\cR_{231}(\pi/6+i\varphi) \\
		\cR_{321}(-\pi/6-i\varphi) \\
		\cR_{312}(\pi/6+i\varphi) \\
		\cR_{132}(-\pi/6-i\varphi) \\
	\end{pmatrix}\overset{S_{2}(-2\pi/3-i\varphi)}{\longrightarrow}
	\begin{pmatrix}
		\cR_{123}(3\pi/2-i\varphi) \\
		\cR_{213}(-3\pi/2+i\varphi) \\
		\cR_{231}(3\pi/2-i\varphi) \\
		\cR_{321}(-3\pi/2+i\varphi) \\
		\cR_{312}(3\pi/2-i\varphi) \\
		\cR_{132}(-3\pi/2+i\varphi) \\
	\end{pmatrix}
\end{equation}
In this way one verifies that the three `forbidden' arguments of Eq.~\eqref{3ParticlePaper_WaveChain} indeed have zero residue. 

With the residues in hand, one constructs the Bethe function in the usual way (c.f. Eq.~\eqref{3ParticlePaper_Aanalytic}). The absence of diffraction means that no disintegration of the bound state occurs, although non-trivial rearrangements are possible. For instance, the outgoing wave at angle $-\pi/6$ in sector 123, corresponding to the argument $-7\pi/6+i\varphi$ in Eq.~\eqref{3ParticlePaper_WaveChain}, describes a bound state of particle 2 and 3, so that the process
\[(1,2)+3\longrightarrow (2,3)+1,\]
is allowed. Note also that reflection of the bound state from the single particle does not occur.

When the Yang--Baxter equation no longer holds, our result Eq.~\eqref{3ParticlePaper_Dsmall} gives the leading order diffraction amplitude describing the disintegration of the bound state upon collision with the other particle.

% \texttt{Three particle bound state is disussed in Ref.~\cite{gaudin1975}}
% \texttt{Earlier work on stripping reaction (Lieb / Koppe )}

\section{Conclusion}
\label{sec:conclusion}

The `quantum Newton's cradle' \cite{kinoshita2006} presumably owes its remarkable features to the distribution function of a 1D gas being unaffected by binary collisions, as described in the Introduction. If an experiment of this type were performed using a gas consisting of more than one species, in which the interaction constants were not all strictly equal, the three body diffractive scattering described in this work will lead to relaxation, which would provide a controlled demonstration of the violation of integrability. In conclusion, we briefly sketch how our result is incorporated into the kinetic description.

Ignoring the possibility of coherence between different particle species (i.e. off diagonal elements of the density matrix), the state of such a gas is described by distribution functions $f_{i}(k,x,t)$, where the index $i$ ranges over the different particle species, and the Boltzmann equation takes the general form
\begin{equation}
	\label{3ParticlePaper_Boltzmann}
	\frac{df_{i}}{dt} = \cI_{\text{coll},i}[\{f_{j}\}].
\end{equation}
Binary collisions \emph{do} contribute to the collision integral on the right hand side, because two particles of different species may bounce of each other, changing their respective $f_{i}$'s. However, the two body collision integral will disappear from the equation for the \emph{total} occupancy $\sum_{i} f_{i}$, which can only be affected by diffractive scattering of three or more particles. 

\begin{figure}
	\centering
	\def\svgwidth{0.8\columnwidth}
	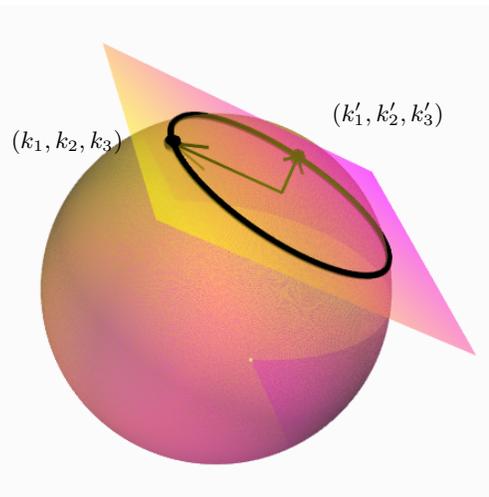
	\caption{Geometry of three particle scattering in momentum space. Allowed values of the three momenta before and after collision lie on the intersection of the sphere of fixed energy and the plane of fixed momentum. The radius of the resulting circle fixes the centre of mass momentum $k$, while the angle between the two points defines $\phi$ and the sector $Q$.}
	\label{fig:ScatteringGeometryCut}
\end{figure}

As an illustration, the three body collision integral describing the change in the occupancy of species 1 due to collisions with species 2 and 3 takes the form (restoring all dimensionful factors)
\begin{multline}
	\label{3ParticlePaper_3CollisionIntegral}
		-\frac{\hbar}{2\pi m}\int \frac{dk_{2}dk_{3}}{(2\pi)^{2}}\sum_{Q}\int_{-\pi/6}^{\pi/6}d\phi\,|\cD_{Q}(\phi,k)|^{2} \\
		\times\left[f_{1}(k_{1})f_{2}(k_{2})f_{3}(k_{3})-f_{1}(k_{1}')f_{2}(k_{2}')f_{3}(k_{3}')\right].
\end{multline}
Eq.~\eqref{3ParticlePaper_3CollisionIntegral} is written for a Boltzmann (non degenerate) gas for brevity, and only the momentum arguments of the distribution functions are shown. To understand the relationship of the angle $\phi$ and wavevector $k$ to the momenta, recall that momentum and energy conservation imply
\begin{equation}
	\begin{split}
		\label{3ParticlePaper_PEfixed}
		k_{1}+k_{2}+k_{3}=k_{1}'+k_{2}'+k_{3}'=P\\
		k_{1}^{2}+k_{2}^{2}+k_{3}^{2}=k_{1}'^{2}+k_{2}'^{2}+k_{3}'^{2}=2mE.		
	\end{split}
\end{equation}
This tells us that in three dimensional momentum space, the allowed values lie on the intersection of the sphere of radius $\sqrt{2mE}$ and the plane parallel to the $(1,1,1)$ direction at a distance $P/\sqrt{3}$ from the origin (Fig.~\ref{fig:ScatteringGeometryCut}). This is a circle of radius $k=\sqrt{2mE-P^{2}/3}$, and the angle $\phi$ and sector $Q$ are determined from the angle between the points $(k_{1},k_{2},k_{3})$ and $(k_{1}',k_{2}',k_{3}')$ on this circle.

The simplest prediction that we can make on this basis is that for a two component Bose gas, the relaxation rate of the overall distribution function will be $\propto n_{1}n_{2}^{2}+n_{2}n_{1}^{2}=(n_{1}+n_{2})n_{1}n_{2}$, where $n_{1,2}$ are the densities of the two components.

We are not aware of many studies of the kinetics driven by three body collisions. Ref.~\cite{lunde2007} investigates the effect of such collisions on transport phenomena in 1D electron systems, but stays within the linearized regime. It would be interesting to seek self-similar solutions of the three particle Boltzmann equation, describing a spatially constant but non-equilibrium distribution function, similar to those found in other circumstances \cite{zakharov1992kolmogorov}.

%\texttt{Singularities in forward scattering presuambly not a problem, because square brackets coincide}

%\bibliography{../Literature/3Part}

\begin{thebibliography}{23}%
\makeatletter
\providecommand \@ifxundefined [1]{%
 \@ifx{#1\undefined}
}%
\providecommand \@ifnum [1]{%
 \ifnum #1\expandafter \@firstoftwo
 \else \expandafter \@secondoftwo
 \fi
}%
\providecommand \@ifx [1]{%
 \ifx #1\expandafter \@firstoftwo
 \else \expandafter \@secondoftwo
 \fi
}%
\providecommand \natexlab [1]{#1}%
\providecommand \enquote  [1]{``#1''}%
\providecommand \bibnamefont  [1]{#1}%
\providecommand \bibfnamefont [1]{#1}%
\providecommand \citenamefont [1]{#1}%
\providecommand \href@noop [0]{\@secondoftwo}%
\providecommand \href [0]{\begingroup \@sanitize@url \@href}%
\providecommand \@href[1]{\@@startlink{#1}\@@href}%
\providecommand \@@href[1]{\endgroup#1\@@endlink}%
\providecommand \@sanitize@url [0]{\catcode `\\12\catcode `\$12\catcode
  `\&12\catcode `\#12\catcode `\^12\catcode `\_12\catcode `\%12\relax}%
\providecommand \@@startlink[1]{}%
\providecommand \@@endlink[0]{}%
\providecommand \url  [0]{\begingroup\@sanitize@url \@url }%
\providecommand \@url [1]{\endgroup\@href {#1}{\urlprefix }}%
\providecommand \urlprefix  [0]{URL }%
\providecommand \Eprint [0]{\href }%
\providecommand \doibase [0]{http://dx.doi.org/}%
\providecommand \selectlanguage [0]{\@gobble}%
\providecommand \bibinfo  [0]{\@secondoftwo}%
\providecommand \bibfield  [0]{\@secondoftwo}%
\providecommand \translation [1]{[#1]}%
\providecommand \BibitemOpen [0]{}%
\providecommand \bibitemStop [0]{}%
\providecommand \bibitemNoStop [0]{.\EOS\space}%
\providecommand \EOS [0]{\spacefactor3000\relax}%
\providecommand \BibitemShut  [1]{\csname bibitem#1\endcsname}%
\let\auto@bib@innerbib\@empty
%</preamble>
\bibitem [{\citenamefont {Arnol'd}(1989)}]{arnold1989}%
  \BibitemOpen
  \bibfield  {author} {\bibinfo {author} {\bibfnamefont {V.}~\bibnamefont
  {Arnol'd}},\ }\href@noop {} {\emph {\bibinfo {title} {Mathematical methods of
  classical mechanics}}},\ Vol.~\bibinfo {volume} {60}\ (\bibinfo  {publisher}
  {Springer},\ \bibinfo {year} {1989})\BibitemShut {NoStop}%
\bibitem [{\citenamefont {Caux}\ and\ \citenamefont {Mossel}(2011)}]{caux2011}%
  \BibitemOpen
  \bibfield  {author} {\bibinfo {author} {\bibfnamefont {J.}~\bibnamefont
  {Caux}}\ and\ \bibinfo {author} {\bibfnamefont {J.}~\bibnamefont {Mossel}},\
  }\href@noop {} {\bibfield  {journal} {\bibinfo  {journal} {Journal of
  Statistical Mechanics: Theory and Experiment}\ }\textbf {\bibinfo {volume}
  {2011}},\ \bibinfo {pages} {P02023} (\bibinfo {year} {2011})}\BibitemShut
  {NoStop}%
\bibitem [{\citenamefont {Sutherland}(2004)}]{sutherland2004}%
  \BibitemOpen
  \bibfield  {author} {\bibinfo {author} {\bibfnamefont {B.}~\bibnamefont
  {Sutherland}},\ }\href@noop {} {\emph {\bibinfo {title} {Beautiful models: 70
  years of exactly solved quantum many-body problems}}}\ (\bibinfo  {publisher}
  {World Scientific Pub Co Inc},\ \bibinfo {year} {2004})\BibitemShut {NoStop}%
\bibitem [{\citenamefont {Kinoshita}\ \emph {et~al.}(2006)\citenamefont
  {Kinoshita}, \citenamefont {Wenger},\ and\ \citenamefont
  {Weiss}}]{kinoshita2006}%
  \BibitemOpen
  \bibfield  {author} {\bibinfo {author} {\bibfnamefont {T.}~\bibnamefont
  {Kinoshita}}, \bibinfo {author} {\bibfnamefont {T.}~\bibnamefont {Wenger}}, \
  and\ \bibinfo {author} {\bibfnamefont {D.}~\bibnamefont {Weiss}},\
  }\href@noop {} {\bibfield  {journal} {\bibinfo  {journal} {Nature}\ }\textbf
  {\bibinfo {volume} {440}},\ \bibinfo {pages} {900} (\bibinfo {year}
  {2006})}\BibitemShut {NoStop}%
\bibitem [{\citenamefont {Mazets}\ and\ \citenamefont
  {Schmiedmayer}(2010)}]{mazets2010}%
  \BibitemOpen
  \bibfield  {author} {\bibinfo {author} {\bibfnamefont {I.}~\bibnamefont
  {Mazets}}\ and\ \bibinfo {author} {\bibfnamefont {J.}~\bibnamefont
  {Schmiedmayer}},\ }\href@noop {} {\bibfield  {journal} {\bibinfo  {journal}
  {New Journal of Physics}\ }\textbf {\bibinfo {volume} {12}},\ \bibinfo
  {pages} {055023} (\bibinfo {year} {2010})}\BibitemShut {NoStop}%
\bibitem [{\citenamefont {Tan}\ \emph {et~al.}(2010)\citenamefont {Tan},
  \citenamefont {Pustilnik},\ and\ \citenamefont {Glazman}}]{tan2010}%
  \BibitemOpen
  \bibfield  {author} {\bibinfo {author} {\bibfnamefont {S.}~\bibnamefont
  {Tan}}, \bibinfo {author} {\bibfnamefont {M.}~\bibnamefont {Pustilnik}}, \
  and\ \bibinfo {author} {\bibfnamefont {L.}~\bibnamefont {Glazman}},\
  }\href@noop {} {\bibfield  {journal} {\bibinfo  {journal} {Physical review
  letters}\ }\textbf {\bibinfo {volume} {105}},\ \bibinfo {pages} {90404}
  (\bibinfo {year} {2010})}\BibitemShut {NoStop}%
\bibitem [{\citenamefont {Olshanii}(1998)}]{olshanii1998}%
  \BibitemOpen
  \bibfield  {author} {\bibinfo {author} {\bibfnamefont {M.}~\bibnamefont
  {Olshanii}},\ }\href@noop {} {\bibfield  {journal} {\bibinfo  {journal}
  {Physical Review Letters}\ }\textbf {\bibinfo {volume} {81}},\ \bibinfo
  {pages} {938} (\bibinfo {year} {1998})}\BibitemShut {NoStop}%
\bibitem [{\citenamefont {Lieb}\ and\ \citenamefont
  {Liniger}(1963)}]{lieb1963}%
  \BibitemOpen
  \bibfield  {author} {\bibinfo {author} {\bibfnamefont {E.}~\bibnamefont
  {Lieb}}\ and\ \bibinfo {author} {\bibfnamefont {W.}~\bibnamefont {Liniger}},\
  }\href@noop {} {\bibfield  {journal} {\bibinfo  {journal} {Physical Review}\
  }\textbf {\bibinfo {volume} {130}},\ \bibinfo {pages} {1605} (\bibinfo {year}
  {1963})}\BibitemShut {NoStop}%
\bibitem [{\citenamefont {McGuire}(1964)}]{mcguire1964}%
  \BibitemOpen
  \bibfield  {author} {\bibinfo {author} {\bibfnamefont {J.}~\bibnamefont
  {McGuire}},\ }\href@noop {} {\bibfield  {journal} {\bibinfo  {journal}
  {Journal of Mathematical Physics}\ }\textbf {\bibinfo {volume} {5}},\
  \bibinfo {pages} {622} (\bibinfo {year} {1964})}\BibitemShut {NoStop}%
\bibitem [{\citenamefont {McGuire}\ and\ \citenamefont
  {Hurst}(1972)}]{mcguire1972}%
  \BibitemOpen
  \bibfield  {author} {\bibinfo {author} {\bibfnamefont {J.}~\bibnamefont
  {McGuire}}\ and\ \bibinfo {author} {\bibfnamefont {C.}~\bibnamefont
  {Hurst}},\ }\href@noop {} {\bibfield  {journal} {\bibinfo  {journal} {Journal
  of Mathematical Physics}\ }\textbf {\bibinfo {volume} {13}},\ \bibinfo
  {pages} {1595} (\bibinfo {year} {1972})}\BibitemShut {NoStop}%
\bibitem [{\citenamefont {Gaudin}\ and\ \citenamefont
  {Derrida}(1975)}]{gaudin1975}%
  \BibitemOpen
  \bibfield  {author} {\bibinfo {author} {\bibfnamefont {M.}~\bibnamefont
  {Gaudin}}\ and\ \bibinfo {author} {\bibfnamefont {B.}~\bibnamefont
  {Derrida}},\ }\href@noop {} {\bibfield  {journal} {\bibinfo  {journal}
  {Journal de Physique}\ }\textbf {\bibinfo {volume} {36}},\ \bibinfo {pages}
  {1183} (\bibinfo {year} {1975})}\BibitemShut {NoStop}%
\bibitem [{\citenamefont {Lipszyc}(1980)}]{lipszyc1980}%
  \BibitemOpen
  \bibfield  {author} {\bibinfo {author} {\bibfnamefont {K.}~\bibnamefont
  {Lipszyc}},\ }\href@noop {} {\bibfield  {journal} {\bibinfo  {journal}
  {Journal of Mathematical Physics}\ }\textbf {\bibinfo {volume} {21}},\
  \bibinfo {pages} {1092} (\bibinfo {year} {1980})}\BibitemShut {NoStop}%
\bibitem [{\citenamefont {McGuire}\ and\ \citenamefont
  {Hurst}(1988)}]{mcguire1988}%
  \BibitemOpen
  \bibfield  {author} {\bibinfo {author} {\bibfnamefont {J.}~\bibnamefont
  {McGuire}}\ and\ \bibinfo {author} {\bibfnamefont {C.}~\bibnamefont
  {Hurst}},\ }\href@noop {} {\bibfield  {journal} {\bibinfo  {journal} {Journal
  of mathematical physics}\ }\textbf {\bibinfo {volume} {29}},\ \bibinfo
  {pages} {155} (\bibinfo {year} {1988})}\BibitemShut {NoStop}%
\bibitem [{\citenamefont {Mehta}\ and\ \citenamefont
  {Shepard}(2005)}]{mehta2005}%
  \BibitemOpen
  \bibfield  {author} {\bibinfo {author} {\bibfnamefont {N.}~\bibnamefont
  {Mehta}}\ and\ \bibinfo {author} {\bibfnamefont {J.}~\bibnamefont
  {Shepard}},\ }\href@noop {} {\bibfield  {journal} {\bibinfo  {journal}
  {Physical Review A}\ }\textbf {\bibinfo {volume} {72}},\ \bibinfo {pages}
  {032728} (\bibinfo {year} {2005})}\BibitemShut {NoStop}%
\bibitem [{\citenamefont {Mehta}\ \emph {et~al.}(2007)\citenamefont {Mehta},
  \citenamefont {Esry},\ and\ \citenamefont {Greene}}]{mehta2007}%
  \BibitemOpen
  \bibfield  {author} {\bibinfo {author} {\bibfnamefont {N.}~\bibnamefont
  {Mehta}}, \bibinfo {author} {\bibfnamefont {B.}~\bibnamefont {Esry}}, \ and\
  \bibinfo {author} {\bibfnamefont {C.}~\bibnamefont {Greene}},\ }\href@noop {}
  {\bibfield  {journal} {\bibinfo  {journal} {Physical Review A}\ }\textbf
  {\bibinfo {volume} {76}},\ \bibinfo {pages} {022711} (\bibinfo {year}
  {2007})}\BibitemShut {NoStop}%
\bibitem [{\citenamefont {Kartavtsev}\ \emph {et~al.}(2009)\citenamefont
  {Kartavtsev}, \citenamefont {Malykh},\ and\ \citenamefont
  {Sofianos}}]{kartavtsev2009}%
  \BibitemOpen
  \bibfield  {author} {\bibinfo {author} {\bibfnamefont {O.}~\bibnamefont
  {Kartavtsev}}, \bibinfo {author} {\bibfnamefont {A.}~\bibnamefont {Malykh}},
  \ and\ \bibinfo {author} {\bibfnamefont {S.}~\bibnamefont {Sofianos}},\
  }\href@noop {} {\bibfield  {journal} {\bibinfo  {journal} {Journal of
  Experimental and Theoretical Physics}\ }\textbf {\bibinfo {volume} {108}},\
  \bibinfo {pages} {365} (\bibinfo {year} {2009})}\BibitemShut {NoStop}%
\bibitem [{\citenamefont {Sommerfeld}(2004)}]{sommerfeld2004}%
  \BibitemOpen
  \bibfield  {author} {\bibinfo {author} {\bibfnamefont {A.}~\bibnamefont
  {Sommerfeld}},\ }\href@noop {} {\bibfield  {journal} {\bibinfo  {journal}
  {Mathematical Theory of Diffraction}\ ,\ \bibinfo {pages} {9}} (\bibinfo
  {year} {2004})}\BibitemShut {NoStop}%
\bibitem [{\citenamefont {Osipov}\ and\ \citenamefont
  {Norris}(1999)}]{osipov1999}%
  \BibitemOpen
  \bibfield  {author} {\bibinfo {author} {\bibfnamefont {A.}~\bibnamefont
  {Osipov}}\ and\ \bibinfo {author} {\bibfnamefont {A.}~\bibnamefont
  {Norris}},\ }\href@noop {} {\bibfield  {journal} {\bibinfo  {journal} {Wave
  motion}\ }\textbf {\bibinfo {volume} {29}},\ \bibinfo {pages} {313} (\bibinfo
  {year} {1999})}\BibitemShut {NoStop}%
\bibitem [{\citenamefont {Norris}\ and\ \citenamefont
  {Osipov}(1999)}]{Norris:1999}%
  \BibitemOpen
  \bibfield  {author} {\bibinfo {author} {\bibfnamefont {A.}~\bibnamefont
  {Norris}}\ and\ \bibinfo {author} {\bibfnamefont {A.}~\bibnamefont
  {Osipov}},\ }\href@noop {} {\bibfield  {journal} {\bibinfo  {journal} {Wave
  motion}\ }\textbf {\bibinfo {volume} {30}},\ \bibinfo {pages} {69} (\bibinfo
  {year} {1999})}\BibitemShut {NoStop}%
\bibitem [{Note1()}]{Note1}%
  \BibitemOpen
  \bibinfo {note} {This slightly frustrating fact is due to the factor
  $e^{-ikr\protect \qopname \relax o{cos}\alpha }$ in Eq.~\protect \textup
  {\hbox {\mathsurround \z@ \protect \normalfont (\ignorespaces \ref
  {3Particle_Sommerfeld}\unskip \@@italiccorr )}}, which differs from the more
  natural choice $e^{ikr\protect \qopname \relax o{cos}\alpha }$, but proves to
  be far more convenient later}\BibitemShut {NoStop}%
\bibitem [{Note2()}]{Note2}%
  \BibitemOpen
  \bibinfo {note} {The minus signs in Eq.~\protect \textup {\hbox
  {\mathsurround \z@ \protect \normalfont (\ignorespaces \ref
  {3ParticlePaper_AllARel}\unskip \@@italiccorr )}} may be at first surprising:
  in Section~\ref {sub:yang_baxter_equation} we will see that it is more
  correctly the \protect \emph {residues} of ${\protect \cal A}_{Q}(\alpha )$
  that are related by the S-matrix}\BibitemShut {NoStop}%
\bibitem [{\citenamefont {Lunde}\ \emph {et~al.}(2007)\citenamefont {Lunde},
  \citenamefont {Flensberg},\ and\ \citenamefont {Glazman}}]{lunde2007}%
  \BibitemOpen
  \bibfield  {author} {\bibinfo {author} {\bibfnamefont {A.}~\bibnamefont
  {Lunde}}, \bibinfo {author} {\bibfnamefont {K.}~\bibnamefont {Flensberg}}, \
  and\ \bibinfo {author} {\bibfnamefont {L.}~\bibnamefont {Glazman}},\
  }\href@noop {} {\bibfield  {journal} {\bibinfo  {journal} {Physical Review
  B}\ }\textbf {\bibinfo {volume} {75}},\ \bibinfo {pages} {245418} (\bibinfo
  {year} {2007})}\BibitemShut {NoStop}%
\bibitem [{\citenamefont {Zakharov}\ \emph {et~al.}(1992)\citenamefont
  {Zakharov}, \citenamefont {L'vov},\ and\ \citenamefont
  {Falkovich}}]{zakharov1992kolmogorov}%
  \BibitemOpen
  \bibfield  {author} {\bibinfo {author} {\bibfnamefont {V.}~\bibnamefont
  {Zakharov}}, \bibinfo {author} {\bibfnamefont {V.}~\bibnamefont {L'vov}}, \
  and\ \bibinfo {author} {\bibfnamefont {G.}~\bibnamefont {Falkovich}},\
  }\href@noop {} {\emph {\bibinfo {title} {Kolmogorov spectra of turbulence 1.
  Wave turbulence.}}},\ Vol.~\bibinfo {volume} {1}\ (\bibinfo  {publisher}
  {Springer, Berlin (Germany)},\ \bibinfo {year} {1992})\BibitemShut {NoStop}%
\end{thebibliography}

%merlin.mbs apsrev4-1.bst 2010-07-25 4.21a (PWD, AO, DPC) hacked
%Control: key (0)
%Control: author (8) initials jnrlst
%Control: editor formatted (1) identically to author
%Control: production of article title (-1) disabled
%Control: page (0) single
%Control: year (1) truncated
%Control: production of eprint (0) enabled
%

\end{document}